\documentclass{article}
\usepackage{authblk}
\usepackage{cprotect}
\usepackage[margin=1in]{geometry}
\usepackage{graphicx}
\graphicspath{{./figures/}}
\usepackage{subcaption}
\usepackage{amssymb}
\usepackage{amsmath}
\usepackage{multirow}
\usepackage{multicol}
\usepackage{authblk}

\usepackage[dvipsnames]{xcolor}
\definecolor{DarkBurntOrange}{RGB}{200,80,0}

\usepackage{hyperref}
\hypersetup{
  colorlinks=true,
  urlcolor=ForestGreen,
  linkcolor=DarkBurntOrange,
  citecolor=blue
}

\usepackage[authoryear,round]{natbib}
\title{\bf Thurston geometries and parameter constraints from SNIa data}
\author[1]{Tanay Gupta \thanks{tanay23@iiserb.ac.in}}
\author[2]{Anshul Verma \thanks{anshulverma.rs.phy19@itbhu.ac.in}}
\author[1]{Sukanta Panda \thanks{sukanta@iiserb.ac.in}}
\author[2]{Pavan K. Aluri \thanks{pavanaluri.phy@itbhu.ac.in}}

\affil[1]{Dept. of Physics, Indian Institute of Science Education \& Research, Bhopal-462066, India}
\affil[2]{Dept. of Physics, Indian Institute of Technology (BHU), Varanasi-221005, India}

\date{}

\begin{document}
\maketitle

\begin{abstract}
Following the numerous evidence for large-scale cosmic isotropy violation with the advent of the `precision cosmology' era, we explore the possible advantages of extending the flat $\Lambda$CDM model to more general models in order to constrain anisotropies in the universe, otherwise absent in the standard model based on FLRW spacetime. Such extensions are offered by the topologically unique Thurston geometries, which are homogeneous but anisotropic spacetime models. In this work, we attempt to distinguish Thurston geometries from one another by introducing anisotropies via different scale factors in different directions, thereby introducing additional model parameters such as shear, eccentricity, curvature, and a preferred axis. We used the latest compilation of Pantheon+ \& SH0ES Type Ia supernova data for deriving model constraints, and found mild evidence of large-scale isotropy violation.
\end{abstract}

\section{Introduction} \label{Sec1}
The standard model, or $\Lambda$CDM model, of cosmology has been remarkably successful in explaining the structure and evolution of the universe. By and large, it very well explains the cosmic microwave background (CMB) anisotropies, the distribution of large-scale structure in the universe, and the observed abundance of light elements, among other phenomena. It also provides a framework for understanding the universe's accelerating expansion through the concept of dark energy, often described by the cosmological constant term ($\Lambda$).

However, the late observations so far of CMB such as that from COBE, WMAP and Planck missions~\cite{bennett1996four, bennett2013nine, aghanim2020planck}, its power on large angular scales \cite{bennett2011seven, akrami2020planck, schwarz2016cmb}, those from type-Ia supernovae (\cite{schwarz2007isotropy, verma2024constraints,colin2011probing, wiltshire2013hubble, appleby2015probing, zhao2019anisotropy, aluri2013anisotropic}), alignment of optical polarization vectors of distant quasars (\cite{hutsemekers1998evidence, hutsemekers2001confirmation, jain2004large, hutsemekers2014alignment}), radio polarization vector alignments (\cite{birch1982universe, jain1999anisotropy, tiwari2013polarization}) and large scale velocity flows being larger than those predicted in $\Lambda$CDM cosmology (\cite{watkins2009consistently}) among others including a very recent study on nearly all-sky catalogs of radio galaxies and quasars on the inconsistency of dipole anisotropy with kinematic aberration and Doppler boosting effects in an FLRW background (\cite{secrest2025colloquium}) compels one to consider much more general cosmological models, without limiting to strict symmetries imposed by the Cosmological Principle (CP).

Motivated thus by the several instances of large-scale statistical isotropy violation and hence in a hope to find any hint for a cosmic preferred axis, in this work we consider spacetime geometries proposed by Thurston (\cite{thurston1982three}  (\cite{d2024thurston, flores2021thurston, smith2025cosmological, novello2020see}) as general extensions of the FLRW model in which the homogeneity is still preserved but there is no such requirement on isotropy, the violation being sourced primarily by the presence of a non-zero \textit{curvature} parameter. The origins of Thurston geometries can be traced back to a very fundamental question: “How does the topology of a surface relate to its geometry and vice versa?” In this context, geometry refers to a Riemannian structure (\cite{awwad2024large}).

Riemannian homogeneous spaces are a fundamental class of manifolds whose study requires techniques from geometry, algebra and group theory. Thurston geometries are a subset of Riemannian homogeneous spaces, famous for Thurston's geometrisation of 3-manifolds, where these appear as fundamental building blocks.


The outlines of this work are as follows. In section \ref{3}, we describe the SNIa dataset that we used for our analysis, write the set of equations so obtained in section \ref{2} in terms of evolution with respect to a dimensionless time variable and the sampling setup that we employed to obtain constraints on our model parameters. In Section \ref{4}, we discuss the results, present an analysis, determine the relevant operating length scales for each geometry based on the constraints obtained, and describe our measure of 'goodness of fit' for the analysis. Finally, in section \ref{5}, we conclude this work.

\section{Thurston spacetimes as world models} \label{2}
\subsection{The geometries} \label{2.1}
Thurston geometries are a set of homogeneous but anisotropic spacetime geometries (\cite{thurston1982three, perelman2002entropy, perelman2003finite, perelman2003ricci}) and are classified as (\cite{awwad2024large}):
\begin{enumerate}
    \item \textbf{3} FLRW spacetimes,
        \begin{equation}
            ds^2 = -c^2 dt^2 + a^2 (t) \{d\chi^2 + S_\kappa^2(\chi) d\Omega^2\}
        \end{equation}
        ($d\Omega^2 = d\theta^2 + \text{sin}^2 \theta d\phi^2$)
        
    \item \textbf{2} FLRW spacetimes in 2D with a third flat anisotropic axis ($\mathbb{R} \times \mathbb{H}^2/S^2$),
        \begin{equation}
            ds^2 = -c^2 dt^2 + a^2 (t) \{dz^2 + d\chi^2 + S_\kappa^2(\chi) d\phi^2\}
        \end{equation}
        \{z $\in$ R is orthogonal to ($\chi$,$\phi$) plane\}\\
        where
        \begin{equation}
            S_\kappa (\chi) =
            \begin{cases}
            \frac{\sin (\chi \sqrt{\kappa} / c)}{(\sqrt{\kappa} / c)}, & \kappa > 0 \hspace{2mm} (S^3, \hspace{1mm} \mathbb{R} \times S^2)\\
            \chi, & \kappa = 0\\
            \frac{\sinh (\chi \sqrt{-\kappa} / c)}{(\sqrt{-\kappa} / c)}, & \kappa < 0 \hspace{2mm} (\mathbb{H}^3, \hspace{1mm} \mathbb{R} \times \mathbb{H}^2)
            \end{cases} 
        \end{equation}   
        where $\kappa$ is the curvature parameter of the universe and is related to the radius of curvature of each Thurston geometry L by
        \begin{equation} \label{kL}
            \kappa = \frac{c^2}{L^2}
        \end{equation}
        where c is the speed of light in vacuum.
        
    \item \textbf{1} Universal cover spacetime of the unit tangent bundle of the hyperbolic plane ($\widetilde{U(\mathbb{H}^2)}$),
        \begin{equation}
            ds^2 = -c^2 dt^2 + a^2(t)\ \left\{dx^2 + \cosh^2 \left(\frac{x\sqrt{-\kappa}}{c}\right)dy^2 + \left(dz + \sinh\left(\frac{x\sqrt{-\kappa}}{c}\right)dy \right)^2\ \right\}
        \end{equation}
        
    \item \textbf{1} Nilpotent spacetime subgroup of an extension of the group of isometries (abb. \textit{Nil}),
        \begin{equation}
            ds^2 = -c^2 dt^2 + a^2 (t) \left\{dx^2 + \left(1 - \frac{\kappa x^2}{c^2}\right)dy^2 + dz^2 - \frac{2x\sqrt{-\kappa}}{c} dydz\right\}
        \end{equation}
        
    \item \textbf{1} Solvable spacetime Lie group (abb. \textit{Solv})
        \begin{equation}
            ds^2 = -c^2 dt^2 + a^2 (t) \{e^\frac{2z\sqrt{-\kappa}}{c}dx^2 + e^{-\frac{2z\sqrt{-\kappa}}{c}}dy^2 + dz^2\}
        \end{equation}
\end{enumerate}
In the next section, we'll attempt to determine the truly anisotropic geometries out of $\mathbb{R} \times \mathbb{H}^2/S^2$, $\widetilde{U(\mathbb{H}^2)}$, Nil \& Solv by introducing anisotropies in the scale factor itself, thus making the shear tensor $\pi^\mu _\nu$ vanish (see \cite{awwad2024large}).

\subsection{Anisotropies in the scale factor} \label{2.2}
We begin with a general spatial ansatz for Thurston geometries
\begin{equation}
    ds^2 = -c^2 dt^2 + \sum^3_{i,j = 1}\gamma_{ij}A_i(t) A_j(t)dx^idx^j
\end{equation}
and equate all off-diagonal terms in the Einstein tensor ($G^\mu_\nu$ $\forall$  $\mu \neq \nu$) to zero, thus obtaining the following constraints on the scale factors, shown in table \ref{Table1} (\cite{awwad2024large})
\begin{table} [h!]
\centering
\begin{tabular}{|c|c|c|}
\hline
\textbf{Spacetime} & \textbf{Constraints} & $\textbf{A}_\textbf{pref}$\\
\hline
$\mathbb{R}^3$ & $\text{A}_1 = \text{A}_2 = \text{A}_3 = \text{a(t)}$, & -\\
& $\kappa=0$ & \\
\hline
$\mathbb{H}^3/ S^3$ & $\text{A}_1 = \text{A}_2 = \text{A}_3 = \text{a(t)}$ & -\\
\hline
$\mathbb{R} \times \mathbb{H}^2/S^2$ & $\text{A}_1 = \text{A}_2 = \text{a(t)}$ & $\text{A}_3 = \text{b(t)}$\\
\hline
$\widetilde{U(\mathbb{H}^2)}$ & $\text{A}_1 = \text{A}_2 = \text{A}_3 = \text{a(t)}$ & -\\
\hline
Nil & $\text{A}_2 = \text{A}_3 = \text{a(t)}$ & $\text{A}_1 = \text{b(t)}$\\
\hline
Solv & $\text{A}_1 = \text{A}_2 = \text{a(t)}$ & $\text{A}_3 = \text{b(t)}$\\
\hline
\end{tabular}
\caption{Anisotropic scale factor constraints and preferred scale factor}
\label{Table1}
\end{table}

Thus, our \textit{truly anisotropic} spacetimes turn out to be
\begin{enumerate}
    \item $\mathbb{R} \times \mathbb{H}^2/S^2$: 
    \begin{equation} \label{Rstar}
        ds^2 = -c^2 dt^2 + a^2(t)[d\chi^2 + S^2_\kappa(\chi)d\phi^2] + b^2(t)dz^2
    \end{equation}
    \item Nil:
    \begin{equation} \label{Nstar}
        ds^2 = -c^2 dt^2 + a^2(t)\left[\left(1 - \frac{\kappa x^2}{c^2}\right)dy^2 + dz^2 - \frac{2x\sqrt{-\kappa}}{c} dydz \right] + b^2(t)dx^2
    \end{equation}
    \item Solv:
    \begin{equation} \label{Sstar}
        ds^2 = -c^2 dt^2 + a^2(t)[e^{\frac{2z\sqrt{-\kappa}}{c}}dx^2 + e^{-\frac{2z\sqrt{-\kappa}}{c}}dy^2] + b^2(t)dz^2
    \end{equation}
\end{enumerate}
where
\begin{equation}
S_\kappa (\chi) =
    \begin{cases}
    \frac{\text{sin}(\chi \sqrt{\kappa} / c)}{(\sqrt{\kappa} / c)}, & \kappa > 0\\
    \chi, & \kappa = 0\\
    \frac{\text{sinh}(\chi \sqrt{-\kappa} / c)}{(\sqrt{-\kappa} / c)}, & \kappa < 0
    \end{cases} 
\end{equation}
In the next section, we will present the redshift distance expressions for these geometries.

\subsection{Redshift distances (z)} \label{2.3}
For Thurston geometries (\eqref{Rstar}-\eqref{Sstar}), we obtain the following redshift expressions (\cite{constantin2023spatially}, see also appendix \ref{appendix:AF}):
\begin{enumerate}
    \item $\mathbb{R} \times \mathbb{H}^2/S^2$ \& Solv
    \begin{equation} \label{red1}
        1 + z = \frac{a(t_0)}{a(t)}\sqrt{\cos^2(\alpha) + \frac{a^2(t)}{b^2(t)}\frac{b^2(t_0)}{a^2(t_0)}\sin^2(\alpha)}
    \end{equation}
    \item Nil
    \begin{equation} \label{red2}
        1 + z = \frac{a(t_0)}{a(t)}\sqrt{\sin^2(\alpha) + \frac{a^2(t)}{b^2(t)}\frac{b^2(t_0)}{a^2(t_0)}\cos^2(\alpha)}
    \end{equation}
\end{enumerate}
where $t_0$ is the present time and $\alpha$ is the angle between the isotropic plane and the direction of SNIa, respectively. Please note that in our derivation in the Solv spacetime, despite the light rays emanating in a curved x-z plane, the final redshift expression is independent of curvature parameter $\kappa$ and is the same as that in $\mathbb{R} \times \mathbb{H}^2/ S^2$ spacetimes in which light rays emanate in a flat $\chi$-z plane.

\subsection{Evolution equations} \label{2.4}
We begin by writing the field equations for our geometries \eqref{Rstar}-\eqref{Sstar} for a perfect matter fluid source
\begin{equation}
    T^\mu _\nu = (-\rho_m c^2, p, p, p)
\end{equation}
where
\begin{equation}
    p = \bar{w} \rho_m c^2
\end{equation}
where $\bar{w}$ is the average equation of state parameter, $\rho_m$ is the matter source fluid, and c is the speed of light in vacuum. Note that $\bar{w}$ accounts only for the perfect matter fluid and differs from the `effective' equation of state parameter that additionally takes into account the effects of the cosmological constant. We will thus determine the cosmological constant energy density parameter later using the constraint equation \eqref{498}.

The continuity equation leads to finding for all the anisotropic geometries,
\begin{equation}
    \Dot{\rho} + 3H(\rho + p) = 0
\end{equation}
where H is the mean Hubble parameter given by
\begin{equation}
    H = \frac{2H_a + H_b}{3}
\end{equation}
We additionally define shear ($\sigma$) as
\begin{equation}
    \sigma = \frac{H_a - H_b}{3H}
\end{equation}
where $H_a = \dot{a}/a$ is the isotropic Hubble parameter and $H_b = \dot{b}/b$ is the anisotropic Hubble parameter.

We obtain the corresponding time evolution differential equations along with a constraint equation for each geometry:
\begin{enumerate}
    \item $\mathbb{R} \times \mathbb{H}^2/S^2$:
    \begin{equation} \label{ce1}
        -\frac{1}{c^2}\left[3H^2(1-\sigma^2) + \frac{\kappa}{a^2}\right] = \frac{8\pi G}{c^4} (-\rho_m c^2) - \Lambda
    \end{equation}

    \item Nil:
    \begin{equation} \label{ce2}
        -\frac{1}{c^2}\left[3H^2(1-\sigma^2) + \frac{\kappa}{4b^2}\right] = \frac{8\pi G}{c^4} (-\rho_m c^2) - \Lambda
    \end{equation}

    \item Solv:
    \begin{equation} \label{ce3}
        -\frac{1}{c^2}\left[3H^2(1-\sigma^2) + \frac{\kappa}{b^2}\right] = \frac{8\pi G}{c^4} (-\rho_m c^2) - \Lambda
    \end{equation}
\end{enumerate}
We further define the mean scale factor (A) and the eccentricity parameter ($e^2$) (characterising deviation from isotropic expansion) as
\begin{equation}
    A = (a^2b)^\frac{1}{3}
\end{equation}
\begin{equation} \label{eccen}
    e^2 = 1 - \frac{b^2}{a^2}
\end{equation}
Note that this definition of mean scale factor can be seen to reconcile easily with our definition of mean Hubble parameter H introduced earlier as
\begin{equation}
    H = \frac{\dot{A}}{A} = \frac{2H_a + H_b}{3}
\end{equation}
Using constraint equations \eqref{ce1} - \eqref{ce3}, it will be convenient to define the dimensionless energy density parameters as
\begin{equation} \label{82}
\begin{split}
    &\frac{8\pi G \rho_m}{3H^2} = \frac{\rho_m}{\rho_\text{Cr}} = \Omega_m\\
    &\frac{\Lambda}{3H^2} = \frac{\rho_\Lambda}{\rho_\text{Cr}} = \Omega_\Lambda
\end{split}
\end{equation}
where the curvature density parameter is given as
\begin{equation} \label{86}
\Omega_\kappa =
\begin{cases}
-\kappa/(3H^2a^2), & \mathbb{R} \times \mathbb{H}^2/S^2 \\
-\kappa/(12H^2b^2), & \text{Nil} \\
-\kappa/(3H^2b^2), & \text{Solv}
\end{cases} 
\end{equation}
Note that equation \eqref{86} is dimensionless as it should be according to our definition of $\kappa$ in equation \eqref{redefinition} (see below). We can thus write the constraint equations \eqref{ce1} - \eqref{ce3} using equations \eqref{82} \& \eqref{86} at present as
\begin{equation} \label{498}
    \boxed{\Omega_{m,0} + \Omega_{\Lambda,0} + \Omega_{\kappa,0} + \sigma^2 _0 = 1}
\end{equation}

\section{Data and Methodology} \label{3}
\subsection{SNe Ia} \label{3.1}
The Pantheon+ SNIa dataset (\cite{verma2024constraints}) is a compilation of 1701 light curves of about 1550 unique spectroscopically confirmed SNIa. The light curves of the rest of the SNIa are either the same SNIa measured by different surveys or are “SN siblings”, meaning SNe found in the same host galaxy. \\To know the distance of a high-redshift SNIa, one requires complete information of the `distance ladder' from the following rungs:
\begin{enumerate}
    \item Cepheid time period and parallax measurements
    \item Independent distance calibration of SNIa-Cephied host galaxies (called \textit{anchors})
    \item High-z (Hubble flow) SNIa observations
\end{enumerate}
The independent host galaxy distances from the II rung, which are calibrated utilising Cepheid (from complementary SH0ES data) period-luminosity information and parallaxes from the first rung, are of utmost importance in terms of removing the degeneracy between the Hubble parameter ‘$H_0$’ and the SNIa absolute magnitude ‘$M_0$’ when performing the cosmological parameter estimation.

\subsection{Evolution equations} \label{3.2}
We define a dimensionless time variable
\begin{equation} \label{455}
    \tau = \text{ln}(A)
\end{equation}
so that
\begin{equation}
    H = \frac{d\tau}{dt}
\end{equation}
We derive (\cite{campanelli2011testing, koivisto2008anisotropic}) for these geometries the direction-dependent luminosity distance ($d_L(\hat{\eta})$) of an SNIa object, observed in the direction $\hat{\eta}$ as
\begin{enumerate}
    \item $\mathbb{R} \times \mathbb{H}^2/S^2$ \& Solv
    \begin{equation} \label{one}
        d_L(\hat{\eta}) = c(1+z_\text{Hel})\int^{1} _{A(z^*)} \frac{dA}{A^2h} \frac{(1-e^2)^\frac{1}{6}}{(1-e^2\cos^2(\alpha))^\frac{1}{2}}
    \end{equation}
    \item Nil
    \begin{equation} \label{two}
         d_L(\hat{\eta}) = c(1+z_\text{Hel})\int^{1} _{A(z^*)} \frac{dA}{A^2h} \frac{(1-e^2)^\frac{1}{6}}{(1-e^2\sin^2(\alpha))^\frac{1}{2}}
    \end{equation}
\end{enumerate}
We'll use the relations between redshift (z) and eccentricity (e) obtained as
\begin{enumerate}
    \item $\mathbb{R} \times \mathbb{H}^2/S^2$ \& Solv
    \begin{equation} \label{59}
        1 + z = \frac{1}{A}\frac{(1 - e^2\cos^2(\alpha))^\frac{1}{2}}{(1 - e^2)^\frac{1}{3}}
    \end{equation}

    \item Nil
    \begin{equation} \label{93}
        1 + z = \frac{1}{A}\frac{(1 - e^2\sin^2(\alpha))^\frac{1}{2}}{(1 - e^2)^\frac{1}{3}}
    \end{equation}
\end{enumerate}
to convert the integration variable from mean scale factor A to the supernova redshift $z^*$. In equations \eqref{one}-\eqref{two}, $z_\text{Hel}$ denotes the heliocentric redshift of an SNIa respectively (\cite{davis2011effect}).

To solve for the integrals in equations \eqref{one}-\eqref{two}, the complete set of common evolution equations in terms of dimensionless variables
\begin{subequations}
\begin{gather}
    H = \frac{2H_a + H_b}{3} = \frac{2\frac{\dot{a}}{a} + \frac{\dot{b}}{b}}{3}\\
    h = \frac{H}{100~\text{km s}^{-1}~ \text{Mpc}^{-1}}\\
    \sigma = \frac{H_a - H_b}{3H} = \frac{\frac{\dot{a}}{a} - \frac{\dot{b}}{b}}{2\frac{\dot{a}}{a} + \frac{\dot{b}}{b}}\\
    r^2 = 1 - \frac{b^2}{a^2}\\
    \Omega_m = \frac{\rho_m}{\rho_{Cr}}\\
    \Omega_\kappa =
    \begin{cases}
        -\kappa/(3H^2a^2), & \mathbb{R} \times \mathbb{H}^2/S^2 \\
        -\kappa/(12H^2b^2), & \text{Nil} \\
        -\kappa/(3H^2b^2), & \text{Solv}
    \end{cases}
\end{gather}
\end{subequations}
can be found using the corresponding field equations to be
\begin{enumerate}
    \item 
    \begin{equation} \label{seventy-five}
        h' = -\frac{3h}{2}\left[1 + \sigma^2 - \frac{\Omega_\kappa}{3} + (1 - \sigma^2 - \Omega_\kappa)\Bar{w}\right]
    \end{equation}

    \item 
    \begin{equation}
        \Omega'_m = -\Omega_m[2\sigma \delta w + 3(\Bar{w}-1)\sigma^2 + \Omega_\kappa(1+3\Bar{w})]
    \end{equation}

    \item 
    \begin{equation}
        (e^2)' = 6\sigma (1-e^2)
    \end{equation}
\end{enumerate}
while the spacetime-specific equations were found to be
\begin{enumerate}
    \item $\mathbb{R} \times \mathbb{H}^2/S^2$
    \begin{enumerate}
        \item 
        \begin{equation} \label{FFirst}
            \sigma' = \frac{(3\sigma \Bar{w} + 2\delta w)(1 - \sigma^2 - \Omega_\kappa)}{2} + (1 + \sigma)(\sigma^2 - \sigma + 1) + \frac{(2 - \sigma)[\Omega_\kappa - (1+\sigma)^2]}{2}
        \end{equation}

        \item 
        \begin{equation}
            \Omega' _\kappa = -2\Omega_\kappa \left(1 + \sigma + \frac{h'}{h}\right)
        \end{equation}
    \end{enumerate}

    \item Nil \& Solv
    \begin{enumerate}
        \item 
        \begin{equation}
            \sigma' = \frac{(3\sigma \Bar{w} + 2\delta w)(1 - \sigma^2 - \Omega_\kappa)}{2} + (1 + \sigma)(\sigma^2 - \sigma + 1) - \frac{(2 - \sigma)[\Omega_\kappa + (1+\sigma)^2]}{2} - \Omega_\kappa (1+\sigma)
        \end{equation}

        \item 
        \begin{equation} \label{eighty-one}
            \Omega' _\kappa = -2\Omega_\kappa \left(1 - 2\sigma + \frac{h'}{h}\right)
        \end{equation}
    \end{enumerate}
\end{enumerate}
where, for some parameter X, its prime denotes a derivative with respect to the new time variable $\tau = \ln{A}$, i.e. 
\begin{equation} \label{tiger}
    X' = \frac{dX}{d\tau}
\end{equation}
One can see that the Thurston geometry \textit{Solv} is a mathematical combination of $\mathbb{R} \times \mathbb{H}^2/S^2$ and Nil, in the sense that it has the same $d_L$ and redshift expressions as that of the former but the same set of evolution equations as the latter. In our analysis, \textit{Solv} was found to be the most computationally expensive for the same precision criteria.

Since we are provided with redshift ($z^*$) of an object, the evolution equations can be solved for any of the cosmological parameters X with respect to redshift z as
\begin{equation}
\begin{split}
    \frac{dX}{dz} &= \frac{dA}{dz} \times \frac{d\tau}{dA} \times \frac{dX}{d\tau}\\
    &= \frac{dA}{dz} \times \frac{1}{A} \times X'
\end{split}
\end{equation}
where dA/dz for each anisotropic geometry can be found directly from equations \eqref{59} \& \eqref{93} as under:
\begin{enumerate}
    \item $\mathbb{R} \times \mathbb{H}^2/S^2$ \& Solv
    \begin{equation} \label{62}
        \frac{dA}{dz} = \frac{A^2(1-e^2)^\frac{1}{3}(1-e^2\cos^2(\alpha))^\frac{1}{2}}{(\sigma+1)e^2 \cos^2(\alpha) + (2-3\cos^2(\alpha))\sigma - 1}
    \end{equation}

    \item Nil
    \begin{equation} \label{79}
        \frac{dA}{dz} = \frac{A^2(1-e^2)^\frac{1}{3}(1-e^2\sin^2(\alpha))^\frac{1}{2}}{(\sigma+1)e^2 \sin^2(\alpha) + (2-3\sin^2(\alpha))\sigma - 1}
    \end{equation}
\end{enumerate}
and again using equations \eqref{59} and \eqref{93} to express A at the RHS in terms of redshift z. Therefore, the evolution equations for each geometry are numerically solved from z = 0 (corresponding to the initial condition $A_0 = 1$ at current time $t = t_0$) up to $z = z^*$ in order to obtain constraints on our model parameters. Finally, the above equation for dA/dz is used to solve our distance integrals in integrals \eqref{one}-\eqref{two} in terms of redshift z (rather than the mean scale factor A).

\subsection{Bayesian setup for deriving cosmological constraints} \label{3.3}
The set of cosmological parameters to be constrained in our models for purely isotropic sources is
\begin{equation} \label{params4}
    \theta = \{\text{h}, \sigma, \text{e}^2, \Omega_m, \Omega_\kappa, \hat{\lambda} = (l_a, b_a), \text{M}_0\, \Bar{w}\}
\end{equation}
where $M_0$ is the absolute B-band magnitude of SNIa. It is assumed that $M_0$ is the same for all SNIa.

Constraints thus obtained on our anisotropic models were compared with the standard flat $\Lambda$CDM model, for which the luminosity distance is given by
\begin{equation} \label{standard}
    d_L (\hat{\eta}) = \frac{c(1+z_\text{Hel})}{h_0} \int^{z^*} _0 \frac{dz}{\sqrt{\Omega_\text{m} (1+z)^3 + \Omega_\Lambda}} 
\end{equation}
The likelihood function is minimised by sampling various parameters that it depends on using the Polychord sampler in Cobaya (\cite{torrado2021cobaya}).

We will be converting the above $\text{d}_\text{L}$ expressions into $\mu^{\text{Th}}$ expressions by using \cite{aluri2013anisotropic, verma2024constraints} as
\begin{equation} \label{mueqn}
    \mu^{\text{Th}} = 5 \, \text{log}_{10}\left(\frac{\text{d}_\text{L}}{10 \, \text{pc}}\right) + \mu_0
\end{equation}
where $\text{d}_\text{L}$ is the dimensionless luminosity distance integral in equations \eqref{one}-\eqref{two} and the constant offset term
\begin{equation}
\begin{split}
    \mu_0 &= 5 \, \text{log}_{10} \left(\frac{c}{100 \, \text{km} \, s^{-1} \, \text{Mpc}^{-1}} \frac{1}{10 \, \text{pc}} \right)\\ 
    &\approx 42.384
\end{split}
\end{equation}
arises as a result of dimensionfull parameters in equations \eqref{one}-\eqref{two} \& equation \eqref{standard} viz., the speed of light c (in km $\text{s}^{-1}$) and the dimensionfull part of the Hubble parameter H = h $\times$ ($100 \, \text{km.s}^{-1} \,\text{Mpc}^{-1}$) and the 10pc term in the denominator of equation \eqref{mueqn}.

We define the $\chi^2$ function as 
\begin{equation} \label{chi2}
    {\chi^2(\theta) = \Delta \mu^T (\text{C}^\text{SN} _\text{stat+sys} + \text{C}^\text{Cep} _\text{stat+sys})^{-1} \Delta \mu}
\end{equation}
In order to obtain constraints, we minimise the $\chi^2(\theta) (= -\text{ln}(L(\theta)))$ function with respect to the parameters $\theta$ that are dependent on our model parameters (equation \eqref{params4}).
There are correlations between SNIa that are calibrated simultaneously due to both statistical and systematic uncertainties in SNIa light curve fitting, for a given Supernova as well as among various SNIa. Therefore, we use the full covariance matrix provided as part of Pantheon+ data release (\cite{scolnic2022pantheon+}).
The priors used for various model parameters in the present work are listed in Table \ref{Table4}.
\begin{table} [h!]
\centering
\begin{tabular}{|c|c|c|}
\hline
Model parameters & \multicolumn{2}{|c|}{Priors}\\
\hline
 & Min & Max\\
\hline
h & 0.5 & 1.0\\
\hline
$\sigma$ & -0.05 & 0.05\\
\hline
$e^2$ & 0 & 0.1\\
\hline
$\Omega_m$ & 0.2 & 0.5\\
\hline
$\Omega_\kappa$ & -0.1 & 0.1\\
\hline
$\bar{w}$ & -1 & 0\\
\hline
$l_a$ & $0^\circ$ & $360^\circ$\\
\hline
$b_a$ & $90^\circ$ & $90^\circ$\\
\hline
$M_0$ & -21 & -18\\
\hline
\end{tabular}
\caption{Priors on various parameters of the model studied in the present work}
\label{Table4}
\end{table}

\section{Results} \label{4}
\subsection{Our models} \label{4.1}
In this section, we present the results from constraining cosmological parameters of Thurston anisotropic models at the current epoch (equations \eqref{one}-\eqref{two}) along with the usual standard cosmological model (equation \eqref{standard}). We use the prior range shown in Table \ref{Table4} for our anisotropic models. Due to improved SNIa calibration techniques (quality) and a significantly larger data sample (quantity) compared to previous releases (\cite{scolnic2022pantheon+, betoule2014improved, suzuki2012hubble, perlmutter1999measurements}), we obtain good constraints on all parameters except the total matter density parameter $\Omega_m$. We will return to this anomaly in section \ref{4.2}.

\begin{figure} [h!]
\begin{subfigure}{.48\textwidth}            
    \centering
    \includegraphics[width=\linewidth]{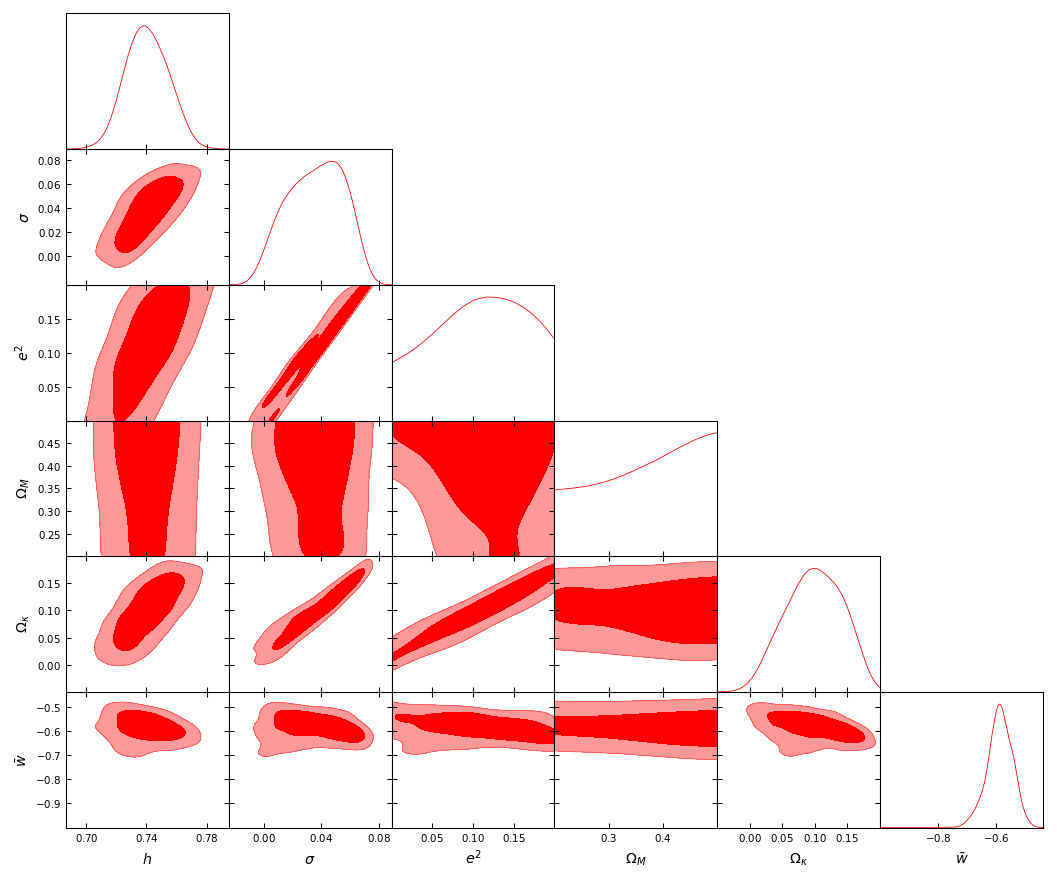}
    \caption{}
    \label{fig:rh2s2_cdm}
\end{subfigure}
\hfill
\begin{subfigure}{.48\textwidth}            
    \centering
    \includegraphics[width=\linewidth]{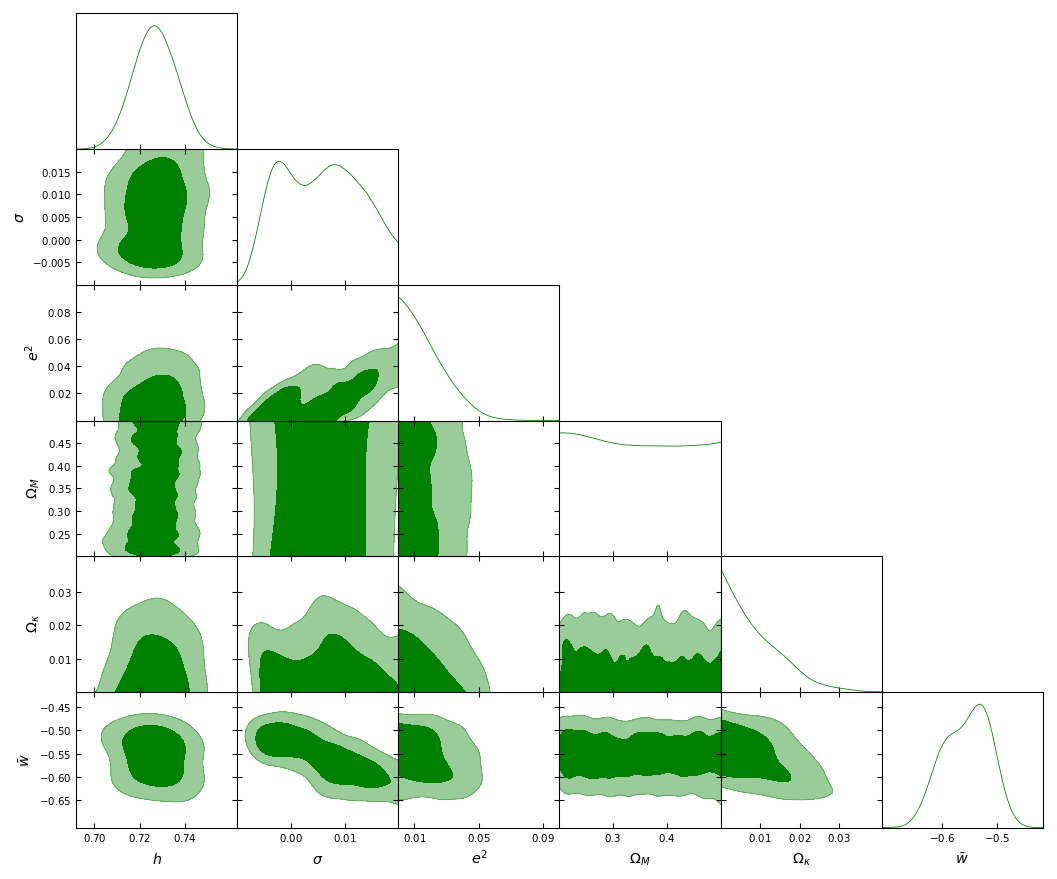}
    \caption{}
    \label{fig:nil_cdm}
\end{subfigure}
\hfill
\begin{subfigure}{.48\textwidth}            
    \centering
    \includegraphics[width=\linewidth]{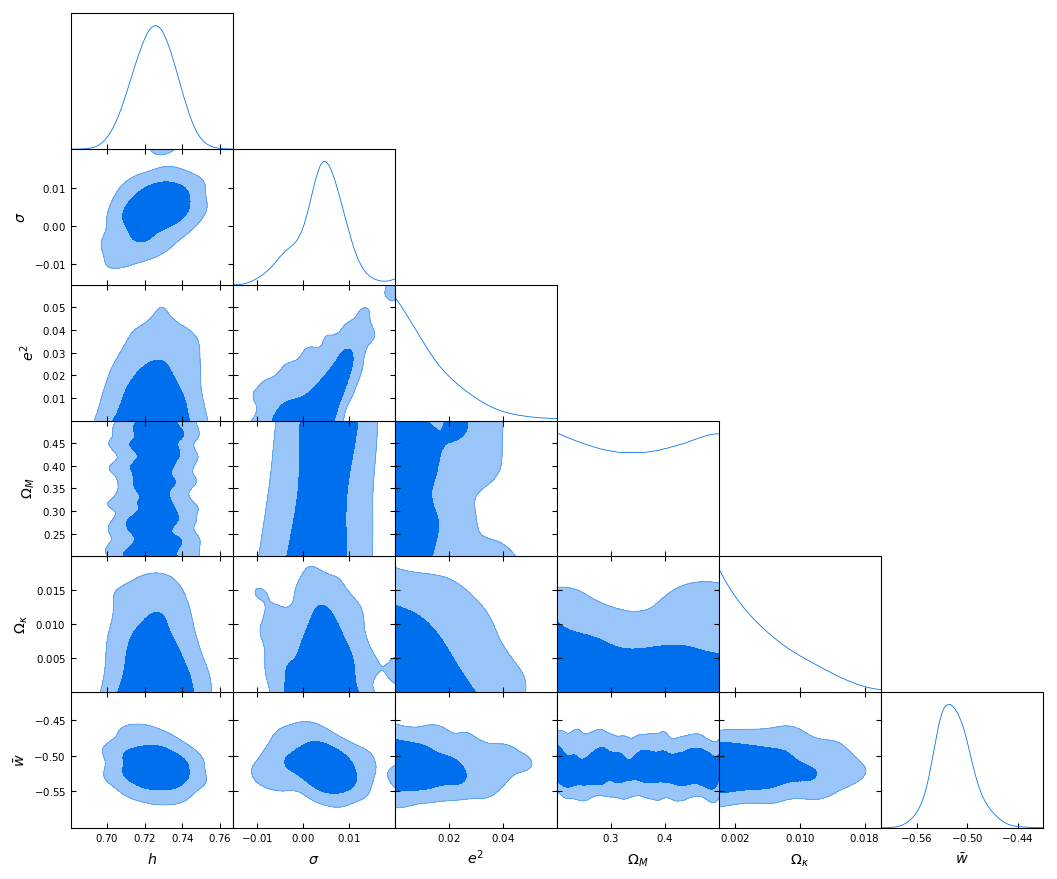}
    \caption{}
    \label{fig:solv_cdm}
\end{subfigure}
\hfill
\begin{subfigure}{.48\textwidth}            
    \centering
    \includegraphics[width=\linewidth]{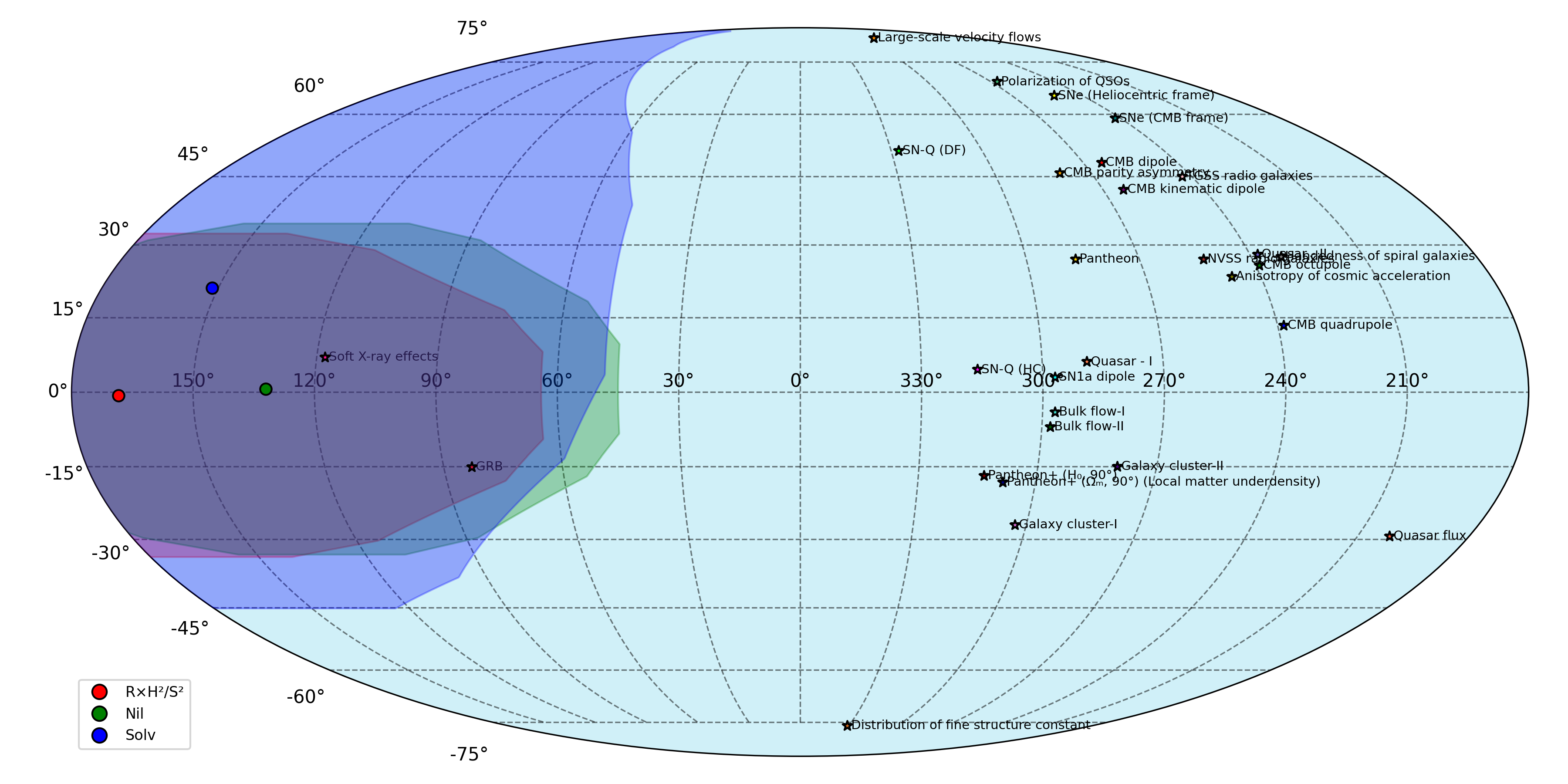}
    \caption{}
    \label{sky_cdm}
\end{subfigure}
\caption{Constraints on geometries and sky map for $\bar{w} = w_a = w_b$}
\label{fig:triangleplots}
\end{figure}

Two-dimensional contour plots with 1$\sigma$ and 2$\sigma$ confidence levels (CLs) for all the cosmological parameters with one anisotropic geometry taken at a time, along with their sky map, are shown in figure \ref{fig:triangleplots}. All the constraints obtained on various model parameters of various anisotropic Thurston geometries, as well as the standard flat $\Lambda$CDM model, are listed in Table \ref{constraints}.

\begingroup
\renewcommand{\arraystretch}{1.25}
\begin{table} [h!]
\centering
\begin{tabular}{|c|c|c|c|c|c|}
\hline
Model & $\Lambda$CDM & $\mathbb{R} \times \mathbb{H}^2/S^2$ & Nil & Solv\\
parameters & & & & \\
\hline
h & 0.733 $\pm$ 0.009 & 0.740 $\pm$ 0.014 & 0.727 $\pm$ 0.010 & 0.725 $\pm$ 0.011\\
\hline
$\sigma$ & - & 0.036 $\pm$ 0.020 & 0.005 $\pm$ 0.007 & 0.004 $\pm$ 0.005\\
\hline
$e^2$ & - & 0.107 $\pm$ 0.053 & 0.018 $\pm$ 0.013 & 0.014 $\pm$ 0.012\\
\hline
$\Omega_M$ & 0.333 $\pm$ 0.018 & 0.368 $\pm$ 0.086 & 0.347 $\pm$ 0.087 & 0.350 $\pm$ 0.089\\
\hline
$\Omega_\kappa$ & - & 0.098 $\pm$ 0.044 & 0.009 $\pm$ 0.007 & 0.006 $\pm$ 0.004\\
\hline
$l_a$ & - & 168.431 $\pm$ 104.449 & 132.048 $\pm$ 87.062 & 151.639 $\pm$ 103.349\\
\hline
$b_a$ & - & -0.702 $\pm$ 33.116 & 0.632 $\pm$ 33.929 & 21.053 $\pm$ 66.260\\
\hline
$M_0$ & -19.252 $\pm$ 0.027 & -19.246 $\pm$ 0.028 & -19.248 $\pm$ 0.028 & -19.245 $\pm$ 0.030\\ 
\hline
$\bar{w}$ & -1 & -0.587 $\pm$ 0.045 & -0.555 $\pm$ 0.043 & -0.516 $\pm$ 0.023\\
\hline
log(Z) & -771.335 $\pm$ 0.305 & -804.051 $\pm$ 0.253 & -805.853 $\pm$ 0.261 & -805.438 $\pm$ 0.269\\ 
\hline
$\Delta$log(Z) & - & 32.716 $\pm$ 0.396 & 34.518 $\pm$ 0.401 & 34.103 $\pm$ 0.407\\
\hline
$\chi^2$ & 1525.19 & 1576.25 & 1577.83 & 1581.75\\
\hline
$L_0/R_0$ & $\infty$ & 1.832 $\pm$ 0.446 & 3.104 $\pm$ 1.229 & 7.609 $\pm$ 2.606\\
\hline
\end{tabular}
\caption{Best fit values for $\Lambda$CDM model \& anisotropic Thurston geometries. Here, $L_0$ implies the curvature radius of each geometry in accordance with \cite{awwad2024large} and `D' the diameter of the Universe.}
\label{constraints}
\end{table}
\endgroup

\subsection{Length scales} \label{4.3}
In this section, we will find the operating length scales for each of our anisotropic Thurston geometries.

We know that the length scale L, which is the radius of curvature of a Thurston geometry, is given by equation \eqref{kL} (\cite{awwad2024large})
\begin{equation} \label{redefinition}
    L = \frac{c}{\sqrt{|\kappa|}}
\end{equation}
where c is the speed of light in vacuum and $\kappa$ is the curvature parameter corresponding to different Thurston geometries.

Using equation \eqref{86}, we have
\begin{equation}
\kappa =
\begin{cases}
-3H^2a^2\Omega_\kappa, & \mathbb{R} \times \mathbb{H}^2/S^2\\
-12H^2b^2\Omega_\kappa, & \text{Nil}\\
-3H^2b^2\Omega_\kappa, & \text{Solv}
\end{cases} 
\end{equation}

so that
\begin{subequations} \label{Ltoday}
\begin{gather}
L_0 = 
\begin{cases} 
\frac{c}{a_0H_0\sqrt{3|\Omega_{\kappa,0}|}}, & \mathbb{R} \times \mathbb{H}^2/S^2\\
\frac{c}{2\sqrt{1 - e^2_0}H_0\sqrt{3\Omega_{\kappa,0}}}, & \text{Nil}\\
\frac{c}{\sqrt{1 - e^2_0}H_0\sqrt{3\Omega_{\kappa,0}}}, & \text{Solv}
\end{cases}
\\
\left|\Delta L_0\right| = 
\begin{cases} 
L_0 \left(\frac{\Delta h_0}{h_0} + \frac{1}{2}\frac{\Delta \Omega_\kappa}{\Omega_\kappa}\right), & \mathbb{R} \times \mathbb{H}^2/S^2\\
\frac{L_0}{2} \left(\frac{\Delta \Omega_\kappa}{\Omega_\kappa} + 2\frac{\Delta h_0}{h_0} - \frac{\Delta e^2_0}{1 - e^2_0}\right), & \text{Nil}\\
\frac{L_0}{2} \left(\frac{\Delta \Omega_\kappa}{\Omega_\kappa} + 2\frac{\Delta h_0}{h_0} - \frac{\Delta e^2_0}{1 - e^2_0}\right), & \text{Solv}
\end{cases}
\end{gather}
\end{subequations}
where we used equation \eqref{eccen} at the present epoch. Using mean and standard deviation values from table \ref{constraints} and the conversion
\begin{equation}
    1 \, \text{km} \, \text{s}^{-1} \, \text{Mpc}^{-1} = 3.24 \times 10^{-20} \, s^{-1}
\end{equation}
We obtain the radii of curvature in meters at present for the above cases as
\begin{equation} \label{radius}
\text{L}_0 =
\begin{cases}
\infty, & \Lambda \text{CDM}\\
(2.308 \pm 0.562) \times 10^{26} \hspace{1mm} \text{m}, & \mathbb{R} \times \mathbb{H}^2/S^2\\
(3.911 \pm 1.549) \times 10^{26} \hspace{1mm} \text{m}, & \text{Nil}\\
(9.587 \pm 3.283) \times 10^{26} \hspace{1mm} \text{m}, & \text{Solv}
\end{cases} 
\end{equation}

Considering the Hubble constant ($H_0$) to be about 73.24 km \, $\text{s}^{-1}$ \, $\text{Mpc}^{-1}$ using HST WF3 anchors NGC 4258, MW \& LMC (\cite{riess20162}), also comparable to the other SNe Ia independent datasets, in particular \cite{suyu2013two}, along with \cite{ade2014planck, sorce2012mid, gao2016megamaser} and \cite{bonamente2006determination}, we obtain the current Hubble radius ($R_0$) as
\begin{equation}
    R_0 = \frac{c}{H_0} = 1.26 \times 10^{26} \, m
\end{equation}
We thus obtain the ratios of curvature radii of each Thurston geometry at present ($L_0$) to that of $R_0$ as
\begin{equation}
\boxed{
\frac{\text{L}_0}{R_0} =
\begin{cases}
\infty, & \Lambda \text{CDM}\\
(1.832 \pm 0.446), & \mathbb{R} \times \mathbb{H}^2/S^2\\
(3.104 \pm 1.229), & \text{Nil}\\
(7.609 \pm 2.606), & \text{Solv}
\end{cases}
}
\end{equation}
thus confirming \cite{awwad2024large}'s claim that the size of individual geometries is much larger than the Hubble radius today.

\subsection{Analysis} \label{4.2}
We note the following observations from our analysis:
\begin{enumerate}
    \item The geometries Nil \& Solv constrain a slightly smaller mean Hubble parameter ($\approx$ 0.72) compared to that of $\Lambda$CDM model ($\approx$ 0.73). 
    
    It suggests that these models might have a slightly slower expansion rate than the $\Lambda$CDM model. This might alleviate some tension between the CMB-derived $H_0$ and local measurements, and could also be interpreted as an effective weakening of late-time dark energy or an extra directional dependence in the expansion, thereby slowing it down.
    
    \item We observe that, out of all the three models, the geometry $\mathbb{R} \times \mathbb{H}^2/S^2$ exhibits the highest mean values of shear ($\approx$ 0.036) and eccentricity ($\approx$ 0.107) respectively. 
    
    This suggests that the geometry $\mathbb{R} \times \mathbb{H}^2/S^2$ exhibits the largest physical deviation from the isotropic standard model.
    
    \item When it comes to constraining the total matter density ($\Omega_m$), all the geometries constrain a higher mean value than that reported by $\Lambda$CDM. This might suggest a slower expansion at earlier times and stronger gravitational clustering in Thurston spacetimes. However, none of the geometries constrained this parameter accurately, with the standard deviations almost five times that of $\Lambda$CDM.
    
    \item Talking about curvature, all our considered geometries constrained fairly small but non-zero values of $\Omega_\kappa$. However, the geometries $\mathbb{R} \times \mathbb{H}^2/S^2$ prefer the \textit{largest} amount of curvature out of all, while \textit{Solv} the \textit{smallest}.

    A foundational observation to note here is that the geometries $\mathbb{R} \times \mathbb{H}^2 / S^2$ prefer a positive value of curvature density parameter $\Omega_\kappa$, which according to equation \eqref{86} directly suggests $\kappa$ to be a negative number (see geometries \eqref{Rstar}-\eqref{Sstar}). Thus, we conclude that the data might prefer $\mathbb{R} \times \mathbb{H}^2$ geometry over $\mathbb{R} \times S^2$.
    
    \item Further, most interestingly, we obtain a common but \textit{unique} set of preferred directions for each of our Thurston geometries, which we compare with the known ones, observing that all the constrained Thurston geometries are in mutual agreement up to 1$\sigma$ for the preferred axis being sourced by soft X-ray effects (\cite{migkas2021cosmological}).
    
    \item  We utilis the Polychord sampler in Cobaya (\cite{torrado2021cobaya}) to compute the Bayesian evidence (log(Z)) of the model under consideration. From the precision-fitting values of log(Z) and minimum $\chi^2$ achieved for all the models, we observe that Pantheon+ SH0ES data still favours the flat $\Lambda$CDM model by a considerable margin. However, among the Thurston geometries, the geometries $\mathbb{R} \times \mathbb{H}^2 / S^2$ is most strongly favoured by the data due to its highest value of log(Z) (after $\Lambda$CDM) and lowest $\chi^2$ (goodness of fit). The difference, however, is negligible, and further analysis might be required to support a concrete claim.
    
    \item Finally, we also estimate the operating length scales for each of our geometries. We observe that while $\Lambda$CDM model has been considered to be perfectly flat (owing to vanishing $\Omega_\kappa$) and capable of describing entire spacetime manifold as a whole, the ratios of curvature radii to the Hubble radii in each Thurston geometry exceeded unity, thus verifying the claim presented in \cite{awwad2024large} that the size of individual geometries is much larger than the Hubble radius today.
\end{enumerate}

\section{Conclusion} \label{5}
Our goal here, assuming Thurston's proposed spacetimes, is to
\begin{itemize}
\item Distinguish between one proposed geometry from the next by considering purely isotropic sources
\item Estimate the level of anisotropy in the observed universe by estimating its cosmic shear and eccentricity 
\item Constrain more general model parameters 
\item Identify any cosmic preferred axis for our universe 
\item Determine the radius of curvature for each anisotropic geometry 
\end{itemize}
We attempt to determine the truly anisotropic geometries from Thurston's proposed ones ($\mathbb{R} \times \mathbb{H}^2/S^2$, Nil \& Solv) by introducing anisotropies in the scale factor and retaining a perfect fluid assumption in the source term in the field equations (\cite{awwad2024large}). The anisotropic models we study here are cosmological models from the \textit{Thurston conjecture}, which classifies all possible geometries (or their combinations, smoothly sewn together) of any given 3D manifold as the large-scale geometries in the observable universe today. We re-derive the evolution equations corresponding to various cosmological parameters of this model in the form of coupled differential equations (equations \eqref{seventy-five}-\eqref{eighty-one}) and perform a Bayesian likelihood analysis of our models, solving these evolution equations in conjunction with the corresponding luminosity distance versus redshift relations (equations \eqref{one} - \eqref{93}) by defining a $\chi^2$ function (equation \eqref{chi2}). For this purpose, we make use of the latest compilation of Type Ia Supernova (SNIa) data from the \textit{Pantheon+} collaboration (\cite{scolnic2022pantheon+}).

Our work extends the analysis presented in (\cite{verma2024constraints}), where constraints on a Bianchi-I universe were obtained using the same dataset as used here, but with anisotropic sources. Starting with uniquely proposed Thurston geometries, we obtain here the constraints on fundamental parameters (Hubble constant $H_0$ \& density parameters $\Omega_\text{m}$ \& $\Omega_\Lambda$), anisotropic ones (eccentricity $e^2$, shear $\sigma$ \& cosmic preferred axis $\hat{\lambda} = (l_a, b_a)$) and geometrical ones (curvature parameter $\Omega_\kappa$) at 2$\sigma$. 

As we assess the evidence for these models to likely describe the data well, we find (from Table \eqref{constraints}) that a flat $\Lambda$CDM model remains preferable. Nevertheless, the models considered here as sources of observed cosmic skewness may still merit further investigation owing to their non-zero constraints at 2$\sigma$ on certain model parameters. 

We therefore conclude from this work that more data may be needed to establish stringent constraints for deciding on Thurston models. In view of this, a succession to this work utilising new \& improved datasets (such as DES, DESI, JWST \& others), based on availability, has a good probability to occur.

\appendix
\section{Redshift distances} 
\label{appendix:AF}
Let us form and solve the geodesic equations in $x^\mu$ (i = 0 to 3) along with one constraining null geodesic equation each for our anisotropic spacetimes, by denoting $dx^\mu/d\lambda = \text{P}^\mu$, where $\lambda$ is the proper geodesic distance.

\subsection{$\mathbb{R} \times \mathbb{H}^2/ S^2$} \label{LooneyTunes}
For $\mathbb{R} \times \mathbb{H}^2$, we have
\begin{enumerate}
    \item \textbf{t equation:}
    \begin{equation}
        \frac{dP^t}{d\lambda} + a\Dot{a}(P^\chi)^2 - \frac{a\Dot{a} \sinh^2(\chi \sqrt{-\kappa}/c)}{(\kappa/c^2)}(P^\phi)^2 + b\Dot{b}(P^z)^2 = 0
    \end{equation}

    \item \textbf{$\chi$ equation:}
    \begin{equation} 
        \frac{dP^\chi}{d\lambda} + 2\frac{\Dot{a}}{a}P^tP^\chi - \frac{\sinh(2\chi \sqrt{-\kappa}/c)}{(2\sqrt{-\kappa}/c)}(P^\phi)^2 = 0
    \end{equation}

    \item \textbf{$\phi$ equation:}
    \begin{equation} 
        \frac{dP^\phi}{d\lambda} + 2\frac{\Dot{a}}{a}P^tP^\phi + \frac{2\sqrt{-\kappa}}{c}\coth(\chi \sqrt{-\kappa}/c)P^\chi P^\phi = 0
    \end{equation}

    \item \textbf{z equation:}
    \begin{equation} 
        \frac{dP^z}{d\lambda} + 2\frac{\Dot{b}}{b}P^t P^z = 0
    \end{equation}

    \item \textbf{Null geodesic constraint:}
    \begin{equation} 
        ds^2 = -c^2 dt^2 + a^2(t)\left[d\chi^2 - \frac{\sinh^2(\chi\sqrt{-\kappa}/c)}{(\kappa/c^2)}d\phi^2\right] + b^2(t)dz^2 = 0
    \end{equation}
\end{enumerate}

while for $\mathbb{R} \times S^2$, we have
\begin{enumerate}
    \item \textbf{t equation:}
    \begin{equation}
        \frac{dP^t}{d\lambda} + a\Dot{a}(P^\chi)^2 + \frac{a\Dot{a}\sin^2(\chi \sqrt{\kappa}/c)}{(\kappa/c^2)}(P^\phi)^2 + b\Dot{b}(P^z)^2 = 0
    \end{equation}

    \item \textbf{$\chi$ equation:}
    \begin{equation} 
        \frac{dP^\chi}{d\lambda} + 2\frac{\Dot{a}}{a}P^tP^\chi - \frac{\sin(2\chi \sqrt{\kappa}/c)}{(2\sqrt{\kappa}/c)}(P^\phi)^2 = 0
    \end{equation}

    \item \textbf{$\phi$ equation:}
    \begin{equation} 
        \frac{dP^\phi}{d\lambda} + 2\frac{\Dot{a}}{a}P^tP^\phi + \frac{2\sqrt{\kappa}}{c}\cot(\chi \sqrt{\kappa}/c)P^\chi P^\phi = 0
    \end{equation}

    \item \textbf{z equation:}
    \begin{equation} \label{qzm}
        \frac{dP^z}{d\lambda} + 2\frac{\Dot{b}}{b}P^t P^z = 0
    \end{equation}

    \item \textbf{Null geodesic constraint:}
    \begin{equation} 
        ds^2 = -c^2 dt^2 + a^2(t)\left[d\chi^2 + \frac{\sin^2(\chi\sqrt{\kappa}/c)}{(\kappa/c^2)}d\phi^2\right] + b^2(t)dz^2 = 0
    \end{equation}
\end{enumerate}
We note that even the brute-forced numerical solution, at least in the case t = ln(a), is highly unlikely to make a(t) scale inversely proportional to (1+z), as expected.

\begin{center}
\begin{figure} [h!]
    \centering
    \includegraphics[width=0.4\linewidth]{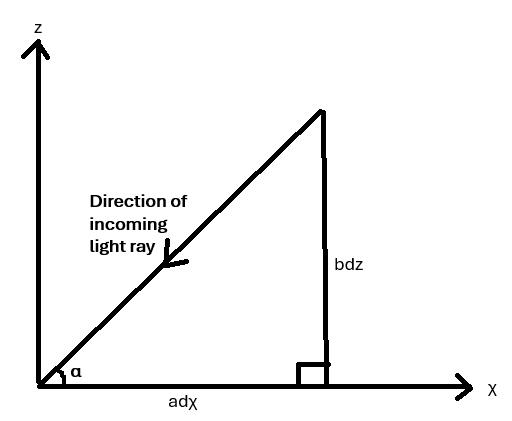}
    \caption{Line of sight in $\mathbb{R} \times \mathbb{H}^2/S^2$ spacetimes}
    \label{fig:rh2s2}
\end{figure}
\end{center}

Considering then 
\begin{equation} 
    P^\chi = v_1a^p(t)
\end{equation}
and
\begin{equation} \label{ppppp}
    P^\phi = v_2a^q(t)
\end{equation}
while we deduce directly from equation \eqref{qzm}
\begin{equation} 
    P^z = \frac{v_3}{b^2(t)}
\end{equation}
for arbitrary positive constants $v_1$, $v_2$ \& $v_3$ with suitable dimensions.\\Then we have
\begin{enumerate}
    \item \textbf{$\chi$ equation:}
    \begin{equation} \label{ccccc}
        (p+2)v_1a^p\Dot{a}P^t - a\frac{\sinh(2\chi \sqrt{-\kappa}/c)}{(2\sqrt{-\kappa}/c)}v^2 _2 a^{2q} = 0
    \end{equation}

    \item \textbf{$\phi$ equation:}
    \begin{equation} \label{zzzzz}
        (q+2)v_2 a^q \Dot{a}P^t + \frac{2\sqrt{-\kappa}}{c}\coth(\chi \sqrt{-\kappa}/c)v_1 v_2 a^{p+q+1} = 0
    \end{equation}
\end{enumerate}
Clearly, the values of p \& q cannot be directly determined from equations \eqref{ccccc} \& \eqref{zzzzz} unless some terms \textit{vanish} (those involving $\kappa$ \& $\chi$) or \textit{blow-up} ($P^t$), which do not appear to be so explicitly.\\
We then use the gauge condition (\cite{constantin2023spatially}) that $\mathbb{R} \times \mathbb{H}^2/S^2$ spacetime reduces to planar Bianchi-I type for source and observer located in $\chi$-z plane, where $\phi$ is constant and hence $P^\phi$ vanishes (see figure \ref{fig:rh2s2}).
We then have, for $\mathbb{R} \times \mathbb{H}^2/S^2$ geometries, in the $\chi$-z plane,
\begin{enumerate}
    \item \textbf{t equation:}
    \begin{equation}
        \frac{dP^t}{d\lambda} + b\Dot{b}(P^z)^2 = 0
    \end{equation}

    \item \textbf{$\chi$ equation:}
    \begin{equation} 
        \frac{dP^\chi}{d\lambda} + 2\frac{\Dot{a}}{a}P^tP^\chi = 0
    \end{equation}

    \item \textbf{z equation:}
    \begin{equation} 
        \frac{dP^z}{d\lambda} + 2\frac{\Dot{b}}{b}P^t P^z = 0
    \end{equation}

    \item \textbf{Null geodesic constraint:}
    \begin{equation} \label{947}
        -c^2 (P^t)^2 + a^2(t)(P^\chi)^2 + b^2(t)(P^z)^2 = 0
    \end{equation}
\end{enumerate}
From figure \ref{fig:rh2s2},
\begin{equation} \label{taneqn}
    tan(\alpha) = \frac{bdz}{ad\chi} = \frac{b}{a}\frac{\frac{dz}{d\lambda}}{\frac{d\chi}{d\lambda}} = \frac{b}{a}\frac{P^z}{P^\chi}
\end{equation}
From the $\chi$ and z equations, we have
\begin{equation} \label{78}
    \boxed{P^\chi = v_1a^{-2}(t)}
\end{equation}
\begin{equation} \label{79}
    \boxed{P^z = v_2b^{-2}(t)}
\end{equation}
and by our gauge condition,
\begin{equation} \label{80}
    \boxed{P^\phi = 0}
\end{equation}
Then
\begin{equation} \label{orange}
    tan(\alpha) = \frac{b(t)}{a(t)} \frac{v_2}{v_1} \frac{a^2(t)}{b^2(t)} = \frac{v_2}{v_1}\frac{a(t)}{b(t)}
\end{equation}
From equations \eqref{947}, \eqref{78} \& \eqref{79}, we obtain
\begin{equation}
    \boxed{P^t = \pm \frac{1}{c}\sqrt{\frac{v_1 ^2}{a^2(t)} + \frac{v_2 ^2}{b^2(t)}}}
\end{equation}
Now, considering the angular frequency $\omega$ has a dimension of velocity  over distance, let us write $\omega$ as:
\begin{equation}
    \omega = \frac{dX^\mu u_\mu}{ds}
\end{equation}
where $u_\mu$ is the four-velocity given by the equation
\begin{center}
    $u_\mu = (c,0,0,0)$
\end{center}
for the rest frame of the spacetime fluid. The sign of $u^\mu$ does not affect the energy-momentum tensor (see appendix B of \cite{weinberg2008cosmology}).

Then
\begin{equation}
    \begin{split}
        \omega &= \frac{dx^0}{d\lambda}u_0 = \frac{dt}{d\lambda}(c)\\
        &= c P^t = \sqrt{\frac{v_1 ^2}{a^2(\tau)} + \frac{v_2 ^2}{b^2(\tau)}}
    \end{split}
\end{equation}
then
\begin{equation}
    \omega(t) = \sqrt{\frac{v_1 ^2}{a^2(t)} + \frac{v_2 ^2}{b^2(t)}}
\end{equation}
and
\begin{equation}
    \omega(t_0) = \sqrt{\frac{v_1 ^2}{a^2(t_0)} + \frac{v_2 ^2}{b^2(t_0)}}
\end{equation}
where $t_0$ is the time parameter at present.

Then, we can write our redshift distance as:
\begin{equation} \label{en103}
    \begin{split}
        1 + z &= \frac{a(t = 0)}{a(t)} = \frac{\lambda(t = 0)}{\lambda(t)} = \frac{\omega(t)}{\omega(t = 0)}\\
        &= \sqrt{\frac{\frac{v^2_1}{a^2(t)}  + \frac{v^2_2}{b^2(t)}}{\frac{v^2_1}{a^2(t_0)}  +  \frac{v^2_2}{b^2(t_0)} }}\\
        &= \sqrt{\frac{\frac{v^2_1}{a^2(t)}}{\frac{v^2_1}{a^2(t_0)} + \frac{v^2_2}{b^2(t_0)}} + \frac{\frac{v^2_2}{b^2(t)}}{\frac{v^2_1}{a^2(t_0)} + \frac{v^2_2}{b^2(t_0)}}}\\
        &= \sqrt{\frac{v^2_1}{a^2(t)} \times \frac{a^2(t_0)b^2(t_0)}{v^2_1 b^2(t_0) + v^2_2 a^2(t_0)} + \frac{v^2_2}{b^2(t)} \times \frac{a^2(t_0)b^2(t_0)}{v^2_1 b^2(t_0) + v^2_2 a^2(t_0)}}\\
    \end{split}
\end{equation}
Now, from equation \eqref{orange}, we have:
\begin{equation}
\begin{split}
    &\cos^2(\alpha) = \frac{v^2_1 b^2(t)}{v^2_2 a^2(t) + v^2_1 b^2(t)}\\
    &\sin^2(\alpha) = \frac{v^2_2 a^2(t)}{v^2_2 a^2(t) + v^2_1 b^2(t)}\\
    &\cos^2(\alpha_0) = \frac{v^2_1 b^2(t_0)}{v^2_2 a^2(t_0) + v^2_1 b^2(t_0)}\\
    &\sin^2(\alpha_0) = \frac{v^2_2 a^2(t_0)}{v^2_2 a^2(t_0) + v^2_1 b^2(t_0)}
\end{split}
\end{equation}
so that equation \eqref{en103} becomes
\begin{equation} \label{zeqn11}
    \begin{split}
        1 + z &= \sqrt{\frac{a^2(t_0)}{a^2(t)} \cos^2(\alpha_0) + \frac{b^2(t_0)}{b^2(t)} \sin^2(\alpha_0)}\\
        &= \frac{a(t_0)}{a(t)}\sqrt{\cos^2(\alpha_0) + \frac{a^2(t)}{b^2(t)}\frac{b^2(t_0)}{a^2(t_0)} \sin^2(\alpha_0)}
    \end{split}
\end{equation}
Here,\\
$a(t_0)$ =  Isotropic scale factor of the universe measured at observation/ present time $t_0$.\\
$a(t)$ =  Isotropic scale factor of the universe measured at time t when light rays were first emitted.\\
$b(t_0)$ =  Anisotropic scale factor of the universe measured at observation/ present time $t_0$.\\
$b(t)$ =  Anisotropic scale factor of the universe measured at time t when light rays were first emitted.\\
$\alpha_0$ = Angle between incoming light ray and $\chi$-axis in $\chi$-z plane at present time.

\subsection{Nil}
For Nil spacetime, we have
\begin{enumerate}
    \item \textbf{t equation:}
    \begin{equation}
        \frac{dP^t}{d\lambda} + b\Dot{b}(P^x)^2 + a\Dot{a}(1-\frac{\kappa x^2}{c^2})(P^y)^2 + a\Dot{a}(P^z)^2 - 2xa\Dot{a}\frac{\sqrt{-\kappa}}{c}P^yP^z = 0
    \end{equation}

    \item \textbf{x equation:}
    \begin{equation} 
        \frac{dP^x}{d\lambda} + 2\frac{\Dot{b}}{b}P^tP^x + \frac{2\kappa x}{c^2} \frac{a^2}{b^2}(P^y)^2 + \frac{\sqrt{-\kappa}}{c}\frac{a^2}{b^2}P^yP^z = 0
    \end{equation}

    \item \textbf{y equation:}
    \begin{equation} 
        \frac{dP^y}{d\lambda} + 2\frac{\Dot{a}}{a}P^tP^y - \frac{\kappa x}{c^2} P^xP^y - \frac{\sqrt{-\kappa}}{c}P^xP^z = 0
    \end{equation}

    \item \textbf{z equation:}
    \begin{equation} 
        \frac{dP^z}{d\lambda} + 2\frac{\Dot{a}}{a}P^tP^z - \frac{\sqrt{-\kappa}}{c}\left(1+\frac{\kappa x^2}{c^2}\right)P^xP^y + \frac{\kappa x}{c^2} P^x P^z = 0
    \end{equation}

    \item \textbf{Null geodesic constraint:}
    \begin{equation} 
        -c^2 (P^t)^2 + a^2(t)\left[\left(1-\frac{\kappa x^2}{c^2}\right)(P^y)^2 + (P^z)^2 - \frac{2x\sqrt{-\kappa}}{c}P^yP^z \right] + b^2(t)(P^x)^2 = 0
    \end{equation}
\end{enumerate}
Again considering
\begin{equation}
    P^x = v_1b^p(t)
\end{equation}
\begin{equation}
    P^y = v_2a^q(t)
\end{equation}
\begin{equation}
    P^z = v_3a^r(t)
\end{equation}
again for arbitrary positive constants $v_1$, $v_2$ \& $v_3$ with suitable dimensions, then
\begin{enumerate}
    \item \textbf{x equation:}
    \begin{equation} 
        (p+2)v_1b^p\Dot{b}P^t + v_2\frac{a^2}{b}\left[2v_2\frac{\kappa x}{c^2} a^{2q} + \frac{\sqrt{-\kappa}}{c}v_3a^{q+r}\right] = 0
    \end{equation}

    \item \textbf{y equation:}
    \begin{equation} 
        (q+2)v_2a^q\Dot{a}P^t - v_1ab^p \left[\frac{\kappa x}{c^2} v_2 a^q + \frac{\sqrt{-\kappa}}{c} v_3a^r \right] = 0
    \end{equation}

    \item \textbf{z equation:}
    \begin{equation} 
        (r+2)v_3a^r\Dot{a}P^t + v_1ab^p \left[\frac{\kappa x}{c^2} v_3 a^r - \frac{\sqrt{-\kappa}}{c}(1+\frac{\kappa x^2}{c^2})v_2 a^q \right] = 0
    \end{equation}
\end{enumerate}
Again, the values of p, q \& r cannot be directly determined from the above equations unless some terms (those involving $\kappa$) vanish or blow up ($P^t$), which do not appear to be so explicitly.

Then, analogous to $\mathbb{R} \times \mathbb{H}^2/S^2$ spacetime, we use the gauge condition that Nil spacetime reduces to planar Bianchi-I type for source and observer located in the x-z plane, where y is constant and hence $P^y$ vanishes.
\begin{center}
\begin{figure} [h!]
    \centering
    \includegraphics[width=0.4\linewidth]{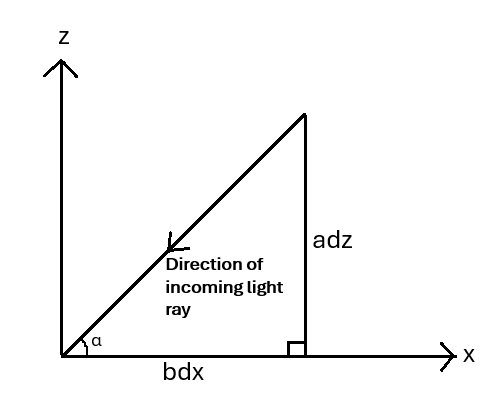}
    \caption{Line of sight in Nil spacetime}
    \label{fig:nil}
\end{figure}
\end{center}
We then have, for Nil geometry, in the x-z plane,
\begin{enumerate}
    \item \textbf{t equation:}
    \begin{equation}
        \frac{dP^t}{d\lambda} + b\Dot{b}(P^x)^2 + a\Dot{a}(P^z)^2 = 0
    \end{equation}

    \item \textbf{x equation:}
    \begin{equation} \label{xeqn}
        \frac{dP^x}{d\lambda} + 2\frac{\Dot{b}}{b}P^tP^x = 0
    \end{equation}

    \item \textbf{z equation:} 
    \begin{equation} \label{yeqn2}
        \frac{dP^z}{d\lambda} + 2\frac{\Dot{a}}{a}P^tP^z = 0
    \end{equation}

    \item \textbf{Null geodesic constraint:}
    \begin{equation} \label{constrt}
        -c^2 (P^t)^2 + a^2(t)(P^z)^2 + b^2(t)(P^x)^2 = 0
    \end{equation}
\end{enumerate}
From figure \ref{fig:nil},
\begin{equation} \label{tan}
    tan(\alpha) = \frac{adz}{bdx} = \frac{a\frac{dz}{d\lambda}}{b\frac{dx}{d\lambda}} = \frac{aP^z}{bP^x}
\end{equation}
From equation \eqref{xeqn}, we have
\begin{equation} \label{V_x}
    \boxed{P^x = v_1b^{-2}(t)}
\end{equation}
From the gauge condition, we have
\begin{equation} \label{V_y}
    \boxed{P^y = 0}
\end{equation}
and from equation \eqref{yeqn2}, we have
\begin{equation} \label{V_Z}
    \boxed{P^z = v_2a^{-2}(t)}
\end{equation}
Substituting values of equations \eqref{V_x} \& \eqref{V_Z} in \eqref{tan}, we get
\begin{equation} \label{Nilsmad}
    \tan(\alpha) = \frac{v_2}{v_1}\frac{b(t)}{a(t)}
\end{equation}
and also into equation \eqref{constrt}, we get
\begin{equation} 
    (P^t)^2 = \frac{v^2 _2}{a^2(t)} + \frac{v^2 _1}{b^2(t)}
\end{equation}
Analogous to the $\mathbb{R} \times \mathbb{H}^2/S^2$ case, we obtain from the null constraint equation,
\begin{equation}
    \omega(t) = cP^t = \sqrt{\frac{v^2 _2}{a^2(t)} + \frac{v^2 _1}{b^2(t)}}
\end{equation}
and
\begin{equation}
    \omega(t_0) = \sqrt{\frac{v^2 _2}{a^2(t_0)} + \frac{v^2 _1}{b^2(t_0)}}
\end{equation}
where $t_0$ is the time parameter at present. Then, we can write our redshift distance as
\begin{equation}
\begin{split}
    1+z &= \frac{a(t = 0)}{a(t)} = \frac{\lambda(t = 0)}{\lambda(t)} = \frac{\omega(t)}{\omega(t = 0)}\\
    &= \sqrt{\frac{\frac{v^2 _2}{a^2(t)} + \frac{v^2 _1}{b^2(t)}}{\frac{v^2 _2}{a^2(t_0)} + \frac{v^2 _1}{b^2(t_0)}}}\\
    &= \sqrt{\frac{\frac{v^2 _2}{a^2(t)}}{\frac{v^2 _2}{a^2(t_0)} + \frac{v^2 _1}{b^2(t_0)}} + \frac{\frac{v^2 _1}{b^2(t)}}{\frac{v^2 _2}{b^2(t_0)} + \frac{v^2 _1}{a^2(t_0)}}}\\
    &= \sqrt{\frac{v^2 _2}{a^2(t)} \times \frac{a^2(t_0)b^2(t_0)}{v^2 _2 b^2(t_0) + v^2 _1 a^2(t_0)} + \frac{v^2 _1}{b^2(t)} \times \frac{a^2(t_0)b^2(t_0)}{v^2 _2 b^2(t_0) + v^2 _1 a^2(t_0)}}\\
\end{split}
\end{equation}
Again, from equation \eqref{Nilsmad}, we have
\begin{equation}
\begin{split}
    &\cos^2(\alpha) = \frac{v^2 _1 a^2(t)}{v^2 _2 b^2(t) + v^2 _1 a^2(\tau)}\\
    &\sin^2(\alpha) = \frac{v^2 _2 b^2(t)}{v^2 _2 b^2(t) + v^2 _1 a^2(\tau)}\\
    &\cos^2(\alpha_0) = \frac{v^2 _1 a^2(t_0)}{v^2 _2 b^2(t_0) + v^2 _1 a^2(\tau_0)}\\
    &\sin^2(\alpha_0) = \frac{v^2 _2 b^2(t_0)}{v^2 _2 b^2(t_0) + v^2 _1 a^2(\tau_0)}
\end{split}
\end{equation}
so that we have
\begin{equation} \label{rednil}
\begin{split}
    1+z &= \sqrt{\frac{a^2(t_0)}{a^2(t)}\sin^2(\alpha_0) + \frac{b^2(t_0)}{b^2(t)}\cos^2(\alpha_0)}\\
    &= \frac{a(t_0)}{a(t)}\sqrt{\sin^2(\alpha_0) + \frac{a^2(t)}{b^2(t)} \frac{b^2(t_0)}{a^2(t_0)} \cos^2(\alpha_0)}
\end{split}
\end{equation}

\subsection{Solv}
For Solv spacetime, we have
\begin{enumerate}
    \item \textbf{t equation:}
    \begin{equation} \label{oneqn3}
        \frac{dP^t}{d\lambda} + a\Dot{a}e^{\frac{2z\sqrt{-\kappa}}{c}}(P^x)^2 + a\Dot{a}e^{\frac{-2z\sqrt{-\kappa}}{c}}(P^y)^2 + b\Dot{b}(P^z)^2 = 0 
    \end{equation}

    \item \textbf{x equation:}
    \begin{equation} 
        \frac{dP^x}{d\lambda} + 2\frac{\Dot{a}}{a}P^tP^x + \frac{2\sqrt{-\kappa}}{c}P^xP^z = 0
    \end{equation}

    \item \textbf{y equation:}
    \begin{equation} 
        \frac{dP^y}{d\lambda} + 2\frac{\Dot{a}}{a}P^tP^y - \frac{2\sqrt{-\kappa}}{c}P^yP^z = 0
    \end{equation}

    \item \textbf{z equation:}
    \begin{equation} 
        \frac{dP^z}{d\lambda} + 2\frac{\Dot{b}}{b}P^tP^z - \frac{\sqrt{-\kappa}}{c}e^{\frac{2z\sqrt{-\kappa}}{c}}\frac{a^2}{b^2}(P^x)^2 + \frac{\sqrt{-\kappa}}{c}e^{\frac{-2z\sqrt{-\kappa}}{c}}\frac{a^2}{b^2}(P^y)^2 = 0
    \end{equation}

    \item \textbf{Null geodesic constraint:}
    \begin{equation} 
        -c^2 (P^t)^2 + a^2(t)[e^{\frac{2z\sqrt{-\kappa}}{c}}(P^x)^2 + e^{\frac{-2z\sqrt{-\kappa}}{c}}(P^y)^2] + b^2(t)(P^z)^2 = 0 
    \end{equation}
\end{enumerate}
Again considering
\begin{equation}
    P^x = v_1a^p(t)
\end{equation}
\begin{equation}
    P^y = v_2a^q(t)
\end{equation}
\begin{equation}
    P^z = v_3b^r(t)
\end{equation}
again for arbitrary positive constants $v_1$, $v_2$ \& $v_3$ with suitable dimensions, then
\begin{enumerate}
    \item \textbf{x equation:}
    \begin{equation} \label{peqn}
        (p+2)\Dot{a}P^t + \frac{2\sqrt{-\kappa}}{c}v_3b^r = 0
    \end{equation}

    \item \textbf{y equation:}
    \begin{equation} \label{qeqb}
        (q+2)\Dot{a}P^t - \frac{2\sqrt{-\kappa}}{c}v_3b^r = 0
    \end{equation}

    \item \textbf{z equation:}
    \begin{equation} \label{reqn}
        (r+2)v_3\Dot{b}b^{r+1}P^t - \frac{\sqrt{-\kappa}}{c} \left[v^2 _1 a^{2p} e^{\frac{2z\sqrt{-\kappa}}{c}} - v^2 _2 a^{2q}e^{\frac{-2z\sqrt{-\kappa}}{c}}\right] a^2 = 0
    \end{equation}
\end{enumerate}
Adding equations \eqref{peqn} and \eqref{qeqb}, we obtain
\begin{equation} \label{kingeqn}
    p + q = -4
\end{equation}
Let us propose the following solutions to equation \eqref{kingeqn}:
\begin{equation}
    p = -2 - \frac{z\sqrt{-\kappa}}{c \cdot \text{ln}(a)}
\end{equation}
\begin{equation}
    q = -2 + \frac{z\sqrt{-\kappa}}{c \cdot \text{ln}(a)}
\end{equation}
so that the second term of equation \eqref{reqn} vanishes in the regimes of late universe a(t) = $e^{At}$ and gauge condition $v_1 = v_2 = v$ and we therefore obtain
\begin{equation}
    r = -2
\end{equation}
Thus, we have
\begin{equation} \label{Solv_1}
    P^x = \frac{v}{a^2(t)}a^{-\frac{z\sqrt{-\kappa}}{c\text{ln}(a)}}
\end{equation}
\begin{equation} \label{Solv_2}
    P^y = \frac{v}{a^2(t)}a^{\frac{z\sqrt{-\kappa}}{c\text{ln}(a)}}
\end{equation}
\begin{equation} \label{Solv_3}
    P^z = \frac{v_3}{b^2(t)}
\end{equation}
We note that, $\forall$ x, p $\in$ N, if
\begin{equation}
   x^\frac{p}{\text{ln x}} = D  
\end{equation}
then upon taking the logarithm of both sides,
\begin{equation}
    \frac{p}{\text{ln x}} \times \text{ln x} = \text{ln D}
\end{equation}
or
\begin{equation}
    \boxed{D = x^\frac{p}{\text{ln x}} = e^p}
\end{equation}
Then equations \eqref{Solv_1}, \eqref{Solv_2} \& \eqref{Solv_3} become
\begin{equation} \label{SolvFina1}
    \boxed{P^x = \frac{v}{a^2(t)}e^{-\frac{z\sqrt{-\kappa}}{c}}}
\end{equation}
\begin{equation} \label{SolvFina2}
    \boxed{P^y = \frac{v}{a^2(t)}e^{\frac{z\sqrt{-\kappa}}{c}}}
\end{equation}
\begin{equation} \label{SolvFina3}
    \boxed{P^z = \frac{v_3}{b^2(t)}}
\end{equation}
Now, since none of the geodesic solutions $P^x, P^y$ \& $P^z$ vanish here, we do not need to stick to any particular plane in the coordinate system, as in previous geometries, to analyse light rays from source to observer.
\begin{center}
\begin{figure} [h!]
    \centering
    \includegraphics[width=0.5\linewidth]{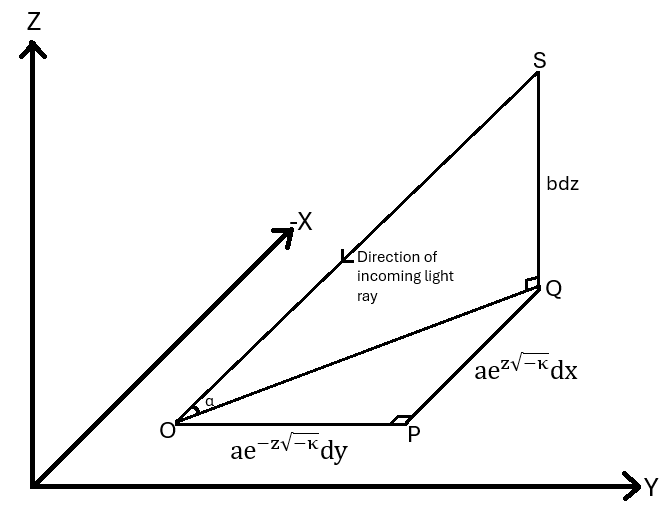}
    \caption{Line of sight in Solv spacetime}
    \label{fig:solvhehe}
\end{figure}
\end{center}
In the ray diagram shown in Figure \ref{fig:solvhehe}, the length OQ is given by
\begin{equation} \label{OQ}
\begin{split}
    OQ &= \sqrt{a^2(t)e^{-\frac{2z\sqrt{-\kappa}}{c}}dy^2 + a^2(t)e^{\frac{2z\sqrt{-\kappa}}{c}}dx^2}\\
   &= a(t)\sqrt{e^{\frac{2z\sqrt{-\kappa}}{c}}dx^2 + e^{-\frac{2z\sqrt{-\kappa}}{c}}dy^2} 
\end{split}
\end{equation}
On dividing equations \eqref{SolvFina1} and \eqref{SolvFina2}, we obtain
\begin{equation}
    \frac{P^x}{P^y} = \frac{dx}{dy} = e^{-\frac{2z\sqrt{-\kappa}}{c}}
\end{equation}
Multiplying both sides by $e^{\frac{z\sqrt{-\kappa}}{c}}$dy, we obtain
\begin{equation}
    e^{\frac{z\sqrt{-\kappa}}{c}}dx = e^{-\frac{z\sqrt{-\kappa}}{c}}dy
\end{equation}
Then equation \eqref{OQ} takes the form
\begin{equation}
    OQ = a(t)\sqrt{e^{\frac{2z\sqrt{-\kappa}}{c}}dx^2 + e^{\frac{2z\sqrt{-\kappa}}{c}}dx^2} = \sqrt{2}a(t)e^{\frac{z\sqrt{-\kappa}}{c}}dx
\end{equation}
Then in $\Delta$SOQ, we have
\begin{equation}
\begin{split}
    \tan(\alpha) &= \frac{QS}{OQ} = \frac{bdz}{\sqrt{2}a(t)e^{\frac{z\sqrt{-\kappa}}{c}}dx}\\
    &= \frac{b(t)}{\sqrt{2}a(t)}\frac{P^z}{P^x}e^{-\frac{z\sqrt{-\kappa}}{c}}
\end{split}
\end{equation}
Again, using equations \eqref{SolvFina1} and \eqref{SolvFina3}, for values of $P^x$ and $P^z$ respectively, we obtain
\begin{equation}
    \tan(\alpha) = \frac{v_3 a(t)}{\sqrt{2}vb(t)}
\end{equation}
so that
\begin{equation} \label{Sine}
    \sin^2(\alpha) = \frac{v^2 _3 a^2(t)}{2v^2b^2(t) + v^2 _3 a^2(t)}
\end{equation}
\begin{equation} \label{Cosine}
    \cos^2(\alpha) = \frac{2v^2b^2(t)}{2v^2b^2(t) + v^2 _3 a^2(t)}
\end{equation}
Again, analogous to the $\mathbb{R} \times \mathbb{H}^2/S^2$ \& Nil spacetimes, we obtain from the null constraint equation,
\begin{equation}
    \omega(t) = cP^t = \sqrt{a^2(t)e^{\frac{2z\sqrt{-\kappa}}{c}}(P^x)^2 + a^2(t)e^{-\frac{2z\sqrt{-\kappa}}{c}}(P^y)^2 + b^2(t)(P^z)^2}
\end{equation}
Substituting values of $P^x$, $P^y$ \& $P^z$ from equations \eqref{SolvFina1} - \eqref{SolvFina3}, this becomes
\begin{equation}
    \omega(t) = \sqrt{\frac{2v^2}{a^2(t)} + \frac{v^2 _3}{b^2(t)}}
\end{equation}
and
\begin{equation}
    \omega(t_0) = \sqrt{\frac{2v^2}{a^2(t_0)} + \frac{v^2 _3}{b^2(t_0)}}
\end{equation}
where $t_0$ is the time parameter at present. Then, we can write our redshift distance as
\begin{equation}
\begin{split}
    1 + z &= \frac{\omega(t)}{\omega(t_0)}\\
    &= \sqrt{\frac{2v^2}{a^2(t)} \times \frac{a^2(t_0)b^2(t_0)}{2v^2b^2(t_0) + v^2 _3 a^2(t_0)} + \frac{v^2 _3}{b^2(t)} \times \frac{a^2(t_0)b^2(t_0)}{2v^2b^2(t_0) + v^2 _3 a^2(t_0)}}\\
    &= \sqrt{\frac{a^2(t_0)}{a^2(t)} \times \frac{2v^2b^2(t_0)}{2v^2b^2(t_0) + v^2 _3 a^2(t_0)} + \frac{b^2(t_0)}{b^2(t)} \times \frac{v^2 _3 a^2(t_0)}{2v^2b^2(t_0) + v^2 _3 a^2(t_0)}}
\end{split}
\end{equation}
Substituting values for $\text{sin}^2(\alpha_0)$ and $\text{cos}^2(\alpha_0)$ from equations \eqref{Sine} and \eqref{Cosine} respectively, this reduces to
\begin{equation}
\begin{split}
    1 + z &= \sqrt{\frac{a^2(t_0)}{a^2(t)}\cos^2(\alpha_0) + \frac{b^2(t_0)}{b^2(t)}\sin^2(\alpha_0)}\\
    &= \frac{a(t_0)}{a(t)}\sqrt{\cos^2(\alpha_0) + \frac{a^2(t)}{b^2(t)}\frac{b^2(t_0)}{a^2(t_0)}\sin^2(\alpha_0)}
\end{split}
\end{equation}

Finally, our redshift distance for various anisotropic spacetimes can be given as
\begin{enumerate}
    \item \textbf{$\mathbb{R} \times \mathbb{H}^2/S^2$:} 
    \begin{equation}
    \begin{split}
        ds^2 &= -c^2 dt^2 + a^2(t)[d\chi^2 + S^2_\kappa(\chi)d\phi^2] + b^2(t)dz^2\\
        &= -c^2 dt^2 + a^2(t)d\chi^2 + b^2(t)dz^2 \hspace{0.1 cm} \text{gauge}
    \end{split}
    \end{equation}

    \begin{equation} \label{321}
        \boxed{1 + z = \frac{a(t_0)}{a(t)}\sqrt{\cos^2(\alpha_0) + \frac{a^2(t)}{b^2(t)}\frac{b^2(t_0)}{a^2(t_0)} \sin^2(\alpha_0)}}
    \end{equation}
    
    \item \textbf{Nil :}
    \begin{equation} 
    \begin{split}
        ds^2 &= -c^2 dt^2 + a^2(t)[(1-\frac{\kappa x^2}{c^2})dy^2 + dz^2 - \frac{2x\sqrt{-\kappa}}{c} dydz] + b^2(t)dx^2\\
        &= -c^2 dt^2 + a^2(t)dz^2 + b^2(t)dx^2 \hspace{0.1 cm} \text{gauge}
    \end{split}
    \end{equation}

    \begin{equation} \label{322}
        \boxed{1 + z = \frac{a(t_0)}{a(t)}\sqrt{\sin^2(\alpha_0) + \frac{a^2(t)}{b^2(t)} \frac{b^2(t_0)}{a^2(t_0)} \cos^2(\alpha_0)}}
    \end{equation}
    
    \item \textbf{Solv :}
    \begin{equation} 
        ds^2 = -c^2 dt^2 + a^2(t)[e^{\frac{2z\sqrt{-\kappa}}{c}}dx^2 + e^{-\frac{2z\sqrt{-\kappa}}{c}}dy^2] + b^2(t)dz^2
    \end{equation}

    \begin{equation} \label{323}
        \boxed{1 + z = \frac{a(t_0)}{a(t)}\sqrt{\cos^2(\alpha_0) + \frac{a^2(t)}{b^2(t)}\frac{b^2(t_0)}{a^2(t_0)}\sin^2(\alpha_0)}}
    \end{equation}
\end{enumerate}
As we can see, in Solv spacetime, despite the light rays emanating in a curved x-z plane, the final redshift expression is independent of curvature parameter $\kappa$ and is the same as that in $\mathbb{R} \times \mathbb{H}^2/ S^2$ spacetimes in which light rays emanate in a flat $\chi$-z plane. The observational distinctions, therefore, between these two geometries would be based on the presence of constraint $P^\phi$ = 0 in $\mathbb{R} \times \mathbb{H}^2/S^2$ spacetime, whereas no such constraint is present in Solv geometry, despite the same redshift distances.

\bibliography{ref}
\end{document}